\documentclass[epj,draft]{svjour}

\begin{document}

\input epsf.sty

\title{Monte Carlo Study of Correlations in Quantum Spin Chains at Non--Zero 
Temperature}

\author{Y. J. Kim\inst{1,2} \and M. Greven\inst{1,} 
\thanks{\emph{Present address:} Department of Applied Physics and Stanford Synchrotron 
Radiation Laboratory, Stanford University, Stanford, CA 94305, USA}
\and U.--J. Wiese\inst{1} \and R. J. Birgeneau\inst{1}} 
\institute{Department of Physics, Massachusetts Institute of 
Technology, Cambridge, MA 02139, USA \and
Division of Engineering and Applied Sciences, Harvard University, 
Cambridge, MA 02138, USA}

\date{Received: date / Revised version: date}

\abstract{
Antiferromagnetic Heisenberg spin chains with various spin values 
($S=1/2,1,3/2,2,5/2$) are studied numerically with the quantum Monte Carlo method.
Effective spin $S$ chains are realized by ferromagnetically coupling 
$n=2S$ antiferromagnetic spin chains with $S=1/2$. 
The temperature dependence of the uniform susceptibility, 
the staggered susceptibility, and the static structure factor peak
intensity are computed down to very low temperatures, $T/J \approx 0.01$. 
The correlation length at each temperature is deduced from 
numerical measurements of the instantaneous spin--spin correlation function. 
At high temperatures, very good 
agreement with exact results for the classical spin chain is obtained 
independent of the value of $S$. 
For the $S$=2 chain which has a gap $\Delta$, the correlation length and the 
uniform susceptibility in the temperature range $\Delta < T < J$ are well 
predicted by the semi--classical theory of Damle and Sachdev.
}
\maketitle

\section{Introduction}

For many years, low--dimensional quantum magnets have drawn much interest 
from both the theoretical and experimental condensed matter physics communities. 
Powerful experimental tools like neutron 
scattering have elucidated the basic physics of 
two--dimensional (2D) antiferromagnets, such as 
$\rm K_2NiF_4$ ($S$=1)\cite{K2NiF4}, 
and one--dimensional (1D) systems, such as $\rm (CD_3)_4NMnCl_3(TMMC)$ ($S$=5/2)
\cite{TMMC,RJB_GS}.
The discovery of high-temperature superconductivity in 1986 
sparked renewed interest in this field, since the parent compounds of 
these superconductors provide very good realizations of 2D 
spin--1/2 square--lattice quantum Heisenberg 
antiferromagnets\cite{Keimer,Greven94}.
Recently, quantum spin ladders and spin chains have also attracted much 
attention, since new copper oxides with such structures have become 
available; these systems are of intrinsic interest and they also allow one to 
compare experimental results with the 
relevant quantum field theories and numerical simulations\cite{Lad_review}.

We report here a Monte Carlo study of antiferromagnetic spin chains as a 
function of spin quantum 
number $S$ and temperature $T$. Recent advances in computer technology as well as the 
development of a powerful loop cluster algorithm\cite{Eve} have 
made it possible to 
perform a very detailed study of various experimentally relevant thermodynamic quantities. 
In this paper, we compute the $S$ and $T$ dependence of the uniform 
susceptibility, $\chi_{\rm u}(S,T)$; the spin--spin correlation 
length, $\xi(S,T)$; the staggered susceptibility, $\chi_{\rm s}(S,T)$; and the static 
structure factor at $q=\pi$, $C_\pi (S,T)$. 
There exists a large 
number of quasi--1D magnetic systems, such as   
$\rm Sr_2CuO_3$ ($S$=1/2) \cite{Ami}, 
copper benzoate ($S$=1/2) \cite{CuBenzo}, 
$\rm Ni(C_2H_8N_2)_2NO_2ClO_4$ ($S$=1) \cite{Regnault}, 
$\rm Y_2BaNiO_5$ ($S$=1) \cite{YBNO},
$\rm CsVCl_3$ ($S$=3/2) \cite{Itoh}, 
$\rm (C_{10}H_8N_2)$$\rm MnCl_3$ ($S$=2) \cite{Granroth}, and 
TMMC ($S$=5/2) \cite{TMMC}.
These magnets all exhibit nearly ideal 1D behavior over a 
considerable range of temperature. Therefore, our study facilitates
comparisons among experimental results, numerical simulations, and theories for such 
materials.

The Hamiltonian for the nearest--neighbor Heisenberg spin chain is 
\begin{equation}
{\cal H} = J \sum_i {\bf S}_i \cdot {\bf S}_{i+1},
\label{hamil}
\end{equation}
where $J$ is positive for an antiferromagnet. We use units in which 
$\hbar=k_B = g\mu_B =1$.
The study of the Hamiltonian Eq.\ (\ref{hamil}) has a long history that dates back to
the remarkable exact solution found by Bethe in 1931 for
$S$=1/2\cite{Bethe}. He found the ground--state eigenfunction for this 
system  and showed that there is no long--range order at $T=0$. 
In 1962, des Cloizeaux and Pearson \cite{Cloiz} derived 
the exact dispersion relation of the lowest--lying excited states at 
$T=0$.
Luther and Peschel\cite{Luther_Pes} showed that the spectrum is gapless without 
magnetic ordering in the ground state, and that the spin correlation function 
decays algebraically with distance. Haldane\cite{Hal83} eventually conjectured that 
half--odd--integer spin chains would behave 
qualitatively like a $S$=1/2 chain,
while for integer spin chains 
the zero--temperature 
spin correlations would decay exponentially with distance
due to the presence of an energy gap, $\Delta$, in the excitation spectrum.
Both numerical and experimental 
confirmation of this conjecture followed\cite{Night,Renard}.
The low--energy and low--temperature properties of integer spin chains are well 
described by the 1D quantum $O(3)$ nonlinear $\sigma$ model 
without any topological term \cite{Hal83}.  
When conformal field theory techniques were applied to 1D 
quantum systems, it was shown that the integrable $S$=1/2 Hamiltonian, at 
low energies, is equivalent to the SU(2) 
Wess-Zumino-Novikov-Witten (WZNW) 
model with a topological coupling constant $k=1$ \cite{WZNW}. 
Moreover, all 
half--odd--integer spin--$S$ Heisenberg models were predicted to be 
equivalent to 
the $k=1$ WZNW model, independent of $S$\cite{Ziman87,RevAff}.

In the limit of classical spins, 
\protect{$S\rightarrow\infty$}, the Hamiltonian Eq.\ (\ref{hamil})
was solved exactly by Fisher in 1964 \cite{Fisher}. 
In order to accommodate the limit $S\rightarrow\infty$,  
Eq.\ (\ref{hamil}) is conveniently rewritten in terms of the unit vector
${\bf \hat{s}}_i \equiv {\bf S}_i / \sqrt{S(S+1)}$, thus introducing 
the energy scale $JS(S+1)$ in place of $J$.
The results for the correlation length and the uniform susceptibility 
per spin are, respectively, 
\begin{equation}
{\xi(S\rightarrow\infty,T) \over a}= - {1 \over \ln u(S,T)}
\label{class_xi}
\end{equation}
and 
\begin{equation}
\chi_{\rm u}(S\rightarrow\infty,T) = {{S(S+1)} \over {3 T}} 
{{1+u(S,T)} \over {1-u(S,T)}},
\label{class_chi}
\end{equation}   
where $a$ is the lattice constant and $u(S,T)$ is given by 
\[
u(S,T)=\coth\left[{{JS(S+1)} \over {T}}\right] - {{T} \over{JS(S+1)}}. 
\]
For the experimental system TMMC ($S=5/2$), the correlation 
lengths in the Heisenberg regime obtained from neutron scattering
could be explained very well by Eq.\ (\ref{class_xi}), without any adjustable parameters, 
since $J$ had been determined independently\cite{TMMC}. 

In Sec.\ \ref{sec:qmc}, we give a brief 
description of our Monte Carlo method. 
Uniform susceptibility data are shown and discussed in Sec.\ 
\ref{sec:unif}, while staggered quantities such as the correlation lengths, 
staggered susceptibilities, and static structure factor peak intensities are 
presented in Sec.\ \ref{sec:stag}. The properties 
of integer spin chains in the context of
recent field theory results are discussed in Sec.\ \ref{sec:disc}. A 
comparison between the results for the spin--$S$ chain and the $n$--chain $S$=1/2 ladder 
is also made in Sec.\ \ref{sec:disc}.

\section{Quantum Monte Carlo}
\label{sec:qmc}

We have carried out quantum Monte Carlo simulations on large lattices 
utilizing the loop cluster algorithm\cite{Eve}.
In order to realize chains with spin $S>1/2$, an $n$--chain 
spin--1/2 ladder
with an infinitely strong ferromagnetic inter--chain coupling
is mapped to a $S=n/2$ chain\cite{Wata}. 
The same algorithm employed to study spin 
ladders is used with minor modifications\cite{Greven}. 
The lengths and Trotter numbers of the chains are chosen so as to minimize any 
finite--size and lattice--spacing effects. 
The chain length is kept at least 10 times larger than the 
calculated correlation length.
Spin states are updated about $10^{4}$ times to reach equilibrium and 
then measured  $10^{5}$ times.
We are able to simulate as many as 100 million lattice points on our workstation.

\section{Uniform Susceptibility}
\label{sec:unif}

\begin{figure}
\centerline{\epsfxsize=3.2in\epsfbox
{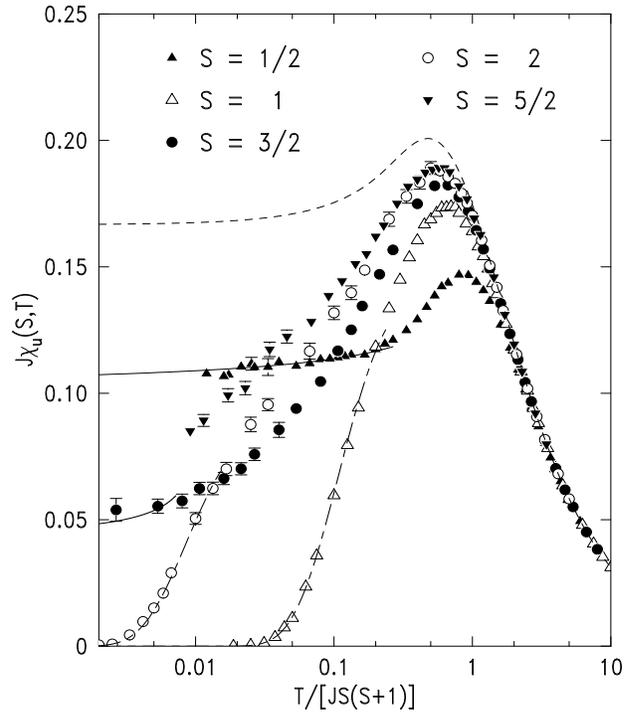}}
\caption{The uniform susceptibility per spin is shown as a function of  
$T/[JS(S+1)]$.  The dashed line is a plot of Fisher's result for the 
classical spin system, Eq.\ 
(\ref{class_chi}). The dot--dashed lines are fits for $S=1$ and $S=2$ to Eq.\ 
(\ref{chi_Tsvelik}).
The solid lines for $S=1/2$ and $S=3/2$ are the WZNW nonlinear $\sigma$ model 
expression, Eq.\ (\ref{chi_lowT}). 
}
\label{unif}
\end{figure}

The uniform susceptibility is shown in Fig.\ 
\ref{unif} 
as a function of  $T/[JS(S+1)]$. 
The dashed line is a 
plot of Fisher's 
result for the classical spin chain, Eq.\ (\ref{class_chi}). At high temperatures, 
$T/[JS(S+1)] > 1$, the 
agreement among the results for all the  
spin values and 
Eq.\ (\ref{class_chi}) is very good. As predicted by Haldane\cite{Hal83}, at
low temperatures the uniform susceptibilities of integer spin chains behave 
markedly differently from 
those of half--odd--integer spin chains. An exponential activation due to 
an  energy gap is 
observed for integer $S$. By fitting the uniform susceptibility 
to the low--temperature expression\cite{Tsvelik,Damle+Sachdev}
\begin{equation}
\chi_{\rm u}(S,T)={1 \over v}
\left({2 \Delta \over \pi T}\right)^{1/2} \exp\left(-{\Delta \over T}\right),
\label{chi_Tsvelik}
\end{equation}
where $\Delta$ is the 
Haldane gap and $v$ is the spin--wave velocity, 
we extract the values $\Delta_{S=1} /J = 
0.40(1)$, $v_{S=1} /Ja=2.5(1)$, $\Delta_{S=2} /J = 0.090(5)$, and $v_{S=2} 
/Ja=4.50(33)$. The uniform susceptibility data are fit only for $T < 
\Delta/2$, and 
the results of this analysis agree with the values deduced in previous 
studies\cite{Night,S2,Sorensen,White}.

For the $S=1/2$ chain, the theoretical low--temperature result for the WZNW 
nonlinear $\sigma$ model \cite{Frischmuth},
\begin{eqnarray}
\chi_{\rm u}(1/2,T)
=&& {1 \over 2\pi v} \nonumber \\ 
+ {1 \over 4\pi v}&&
\left[
{1 \over \ln(T_0/T)} - {\ln(\ln(T_0/T)+1/2) \over 2\ln(T_0/T)^2}
\right],   
\label{chi_lowT}
\end{eqnarray}
is plotted as a solid line.
Here the spin--wave velocity $v_{S=1/2}/Ja=\pi/2$, and $T_0/J\approx1.8$.
The hypothesis that all half--odd--integer Heisenberg chains are in the same 
universality 
class as the $S=1/2$ chain, namely the $k=1$ WZNW model, is now 
generally accepted. However, this claim was initially considered to be 
controversial, since integrable spin-$S$ Hamiltonians are equivalent to the 
WZNW model with $k=2S$\cite{Schulz}. 
A recent numerical study by Hallberg et al. \cite{Hallberg} gave strong evidence 
that $k=1$ for the WZNW model of the $S=3/2$ chain, thus supporting the above 
hypothesis.  
We give here additional evidence in favor of this claim. The asymptotic value of the 
uniform susceptibility for the $k=2S$ WZNW model is different from that of Eq.\ 
(\ref{chi_lowT}). 
Specifically, at $T \rightarrow 0$, $\chi_{\rm u}(S,T) \rightarrow S/\pi v_S$ rather than 
$\chi_{\rm u} (S,0)= 1/2\pi v_S$. If we use the calculated value $v_{S=3/2}/Ja=3.87$
for the spin--wave velocity\cite{Hallberg}, we 
obtain $J\chi_{\rm u}(3/2,0) \approx 0.12$ for the $k=2S$ model and $J\chi_{\rm u}(3/2,0) 
\approx 0.04$ for the $k=1$ model.
Therefore, as is evident from Fig.\ \ref{unif}, our data for $S=3/2$ show 
clearly that the $S=3/2$ Heisenberg chain is equivalent to the WZNW model with $k=1$.
We have, in fact, fitted our $S=3/2$ chain data for $\chi_{\rm u}$ with Eq.\ (\ref{chi_lowT}).
The spin--wave velocity is held fixed at the value given above and only $T_0$ is 
adjusted. 
As may be seen in Fig.\ \ref{unif} the fit is very good at low 
temperatures, as 
expected. For the $S=5/2$ chain, we are not able to obtain numerical data down to low 
enough temperatures to yield a meaningful test of Eq.\ (\ref{chi_lowT}).

\section{Correlation Length, Staggered Susceptibility, and Static Structure 
Factor Peak Intensity}
\label{sec:stag}

\begin{figure}
\centerline{\epsfxsize=3.2in\epsfbox
{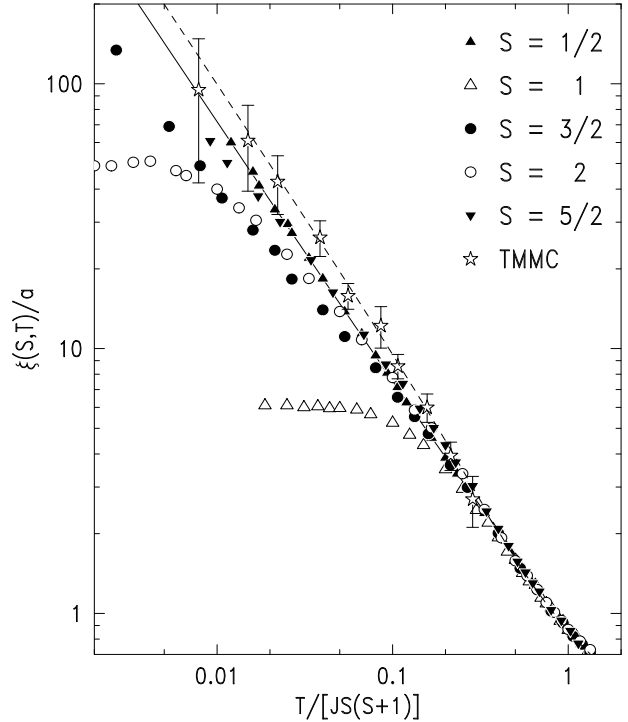}}
\caption{The spin-spin correlation length deduced from fitting the 
computed correlation function to the asymptotic form, Eq.\ (\ref{OZ}) is 
plotted as a function of temperature. 
Also shown are the correlation lengths of 
TMMC from Ref.\ \cite{TMMC}. 
The classical spin system result, Eq.\ (\ref{class_xi}), is shown  as a 
dashed line. The solid line is Eq.\ (\ref{xi_WZNW}) for the $S=1/2$ chain.
}
\label{xi}
\end{figure}

In order to deduce the spin--spin correlation length, $\xi(S,T)$, the 
instantaneous spin--spin 
correlation function, $C(r)$, is computed and fitted to the asymptotic form 
\begin{equation}
C(r) \sim { \exp(-r/\xi) \over r^\lambda}.
\label{OZ}
\end{equation}
Equation\ (\ref{OZ}) is equivalent to 1D and 2D Ornstein-Zernike (OZ) forms when 
$\lambda=0$ and 
$\lambda=1/2$, respectively. Only data with $r > 3\xi$ are included in the fit to 
ensure that the asymptotic behavior is probed. 
For half--odd--integer spin chains, the 1D OZ form is found 
to work very well over the entire temperature range. 
The spin correlations for integer spin chains, however, are found to go through a 
crossover with decreasing temperature, 
from the 1D to the 2D OZ form at $T\approx\Delta/2$. Specifically, 
the 2D OZ form gives a better description of the computed spin correlations at 
low temperatures. This behavior also was observed previously 
for spin ladders of even width\cite{Greven}.
The low--temperature correlation function for the $S=1$ chain is known to be proportional
to the modified Bessel function $K_0 (r/\xi)$\cite{Sorensen}; $K_0$ is also used to 
fit our integer spin chain data, and the results are consistent with 
those extracted using Eq.\ (\ref{OZ}).

In Fig.\ \ref{xi},  the numerical correlation length is shown together with 
$\xi(T)$ for TMMC obtained from neutron scattering 
experiments\cite{TMMC}. 
The result for the classical spin system, Eq.\ (\ref{class_xi}), is shown  as 
the dashed line.
The agreement among the spin correlation lengths for the different quantum 
spin chains, and with the classical curve, is very good for temperatures as low as 
$T/[JS(S+1)] \approx 0.2$, which corresponds to $\xi/a \approx 4$. 
As the temperature is lowered further, the data begin to deviate from 
the classical curve. Note, that for $S \geq 1$, as the spin value is increased,
the agreement 
between the quantum and classical results persists down to progressively lower 
temperatures. Because of the 
presence of a gap for integer spin chains, $\xi(T)$ remains finite as 
$T\rightarrow0$. We 
estimate that $\xi(1,0)/a = 6.0(1)$ and $\xi(2,0)/a = 50(1)$.
These results satisfy the relation $v=\Delta \xi$ with the values for $v$ and 
$\Delta$ obtained above; $\Delta_{S=1} /J = 
0.40(1)$, $v_{S=1} /Ja=2.5(1)$, $\Delta_{S=2} /J = 0.090(5)$, and $v_{S=2} 
/Ja=4.50(33)$, and also agree 
with those of previous numerical studies within the combined 
errors \cite{White,Nomura89}.
For the $S=1/2$ chain, we also plot in Fig.\ \ref{xi} the 
WZNW model prediction, 
\begin{equation}
{1 \over \xi(1/2,T)}=T
\left[2-
{1 \over \ln(T_0/T)} + {\ln(\ln(T_0/T)+1/2) \over 2\ln(T_0/T)^2}
\right],   
\label{xi_WZNW}
\end{equation}
with the parameters obtained by Nomura and Yamada in a 
thermal Bethe ansatz study \cite{Nomura91}.


\begin{figure}
\centerline{\epsfxsize=3.2in\epsfbox
{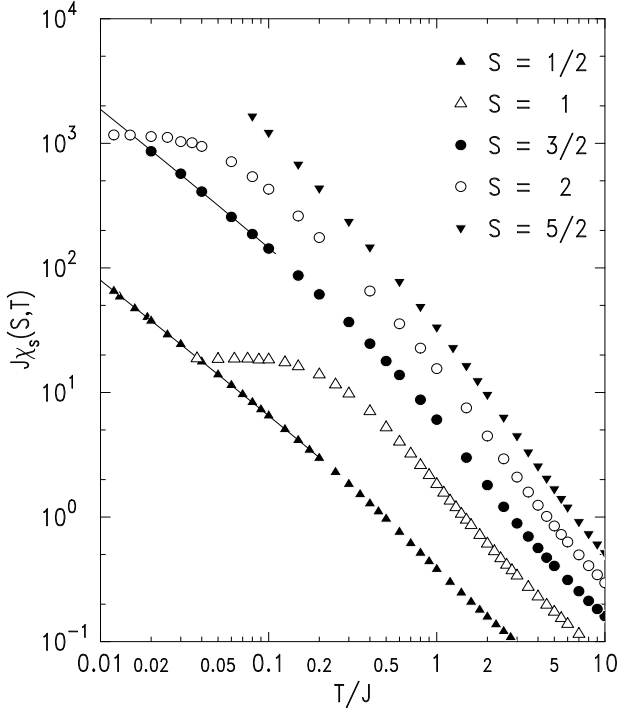}}
\caption{The temperature dependence of the staggered susceptibility per spin.
The solid lines are the results of fits to  Eq.\ 
(\ref{starykh_eq:a}). }
\label{stag}
\end{figure}

\begin{figure}
\centerline{\epsfxsize=3.2in\epsfbox
{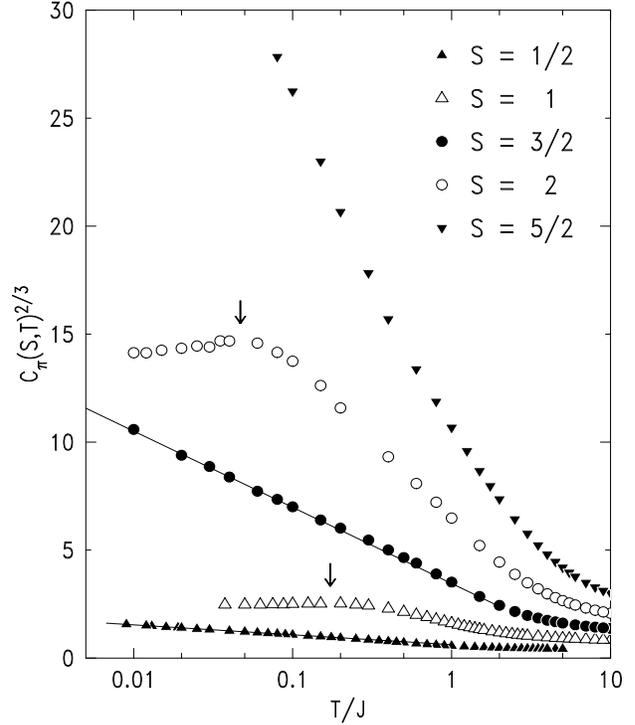}}
\caption{Static structure factor peak intensity $C_\pi(S,T)$ 
is plotted to show the linear 
dependence of $C_\pi^{2/3}$ on $\log T$. The solid lines are fits to 
Eq.\ (\ref{starykh_eq:b}). The arrows indicate the peak positions as discussed in the text.
}
\label{sst}
\end{figure}

The staggered susceptibility per spin is shown in Fig.\ 
\ref{stag}. 
The different behaviors of integer and half--odd--integer spin chains are 
clearly manifest 
in this plot. In Fig.\ \ref{sst}, the static structure factor $C_q$ at $q=\pi$ is plotted 
as a 
function of temperature. Note, that for integer spin chains $C_\pi (S,T)$ 
peaks at $T\approx\Delta/2$; closely similar behavior is observed for spin-1/2 
ladders of even width \cite{Greven}. The extrapolated zero--temperature values 
for the $S=1$ 
chain, $\chi_{\rm s}(1,0)=18.6(1)$ and $C_\pi (1,0)=3.83(2)$ agree  well  with the 
corresponding results of the exact diagonalization study by Sakai and 
Takahashi \cite{Sakai}. 
We also obtained $\chi_{\rm s}(2,0)=1160(10)$, and $C_\pi(2,0)=52.0(3)$ for the $S=2$ 
chain.

Recently, Starykh, Sandvik, and Singh \cite{Starykh} studied the static structure 
factor and the staggered susceptibility of the $S=1/2$ chain and obtained
low--temperature analytic forms for  these quantities:
\begin{equation}
\chi_{\rm s}(S,T) = D_\chi(S) T^{-1} \left[\ln(T_\chi (S) / T) \right]^{1/2},
\label{starykh_eq:a}
\end{equation}
\begin{equation}
C_\pi (S,T) = D_{\rm s} (S) \left[\ln(T_{\rm s} (S) / T) \right]^{3/2}.
\label{starykh_eq:b}
\end{equation} 
We are able to fit our $S=1/2$ data to these forms and thereby to extract 
$D_\chi(1/2)=0.30(1)$, $T_\chi(1/2)=9.8(1.2)$, $D_{\rm s}(1/2)=0.091(1)$, 
and $T_{\rm s}(1/2)=21(1)$.
The solid lines in Figs.\ \ref{stag} and \ref{sst} are the results of 
fits to Eqs.\ (\ref{starykh_eq:a}) and (\ref{starykh_eq:b}) for $T/J < 0.25$.
The parameters thus obtained agree with those of Ref.\ \cite{Starykh} 
to within 5\%, except for $T_\chi(1/2)$. This is due to the difference in 
fitting range, since only 
very low temperature data show asymptotic behavior for $\chi_{\rm s}(S,T)$. 
Our fitting range is $0.009 \leq T/J < 0.25$, while $0.035 \leq T/J < 0.25$ 
was used in Ref.\ \cite{Starykh}.
Our $S=3/2$ data can also be fitted to Eqs.\ (\ref{starykh_eq:a}) and 
(\ref{starykh_eq:b}) for  $T/J \leq 0.1$. The 
parameters so--obtained are $D_\chi(3/2)=7.8(3)$, $T_\chi(3/2)=3.2(1.1)$, $D_{\rm 
s}(3/2)=0.63(2)$, and $T_{\rm s}(3/2)=9.5(1.2)$. 

\begin{figure}
\centerline{\epsfxsize=3.2in\epsfbox
{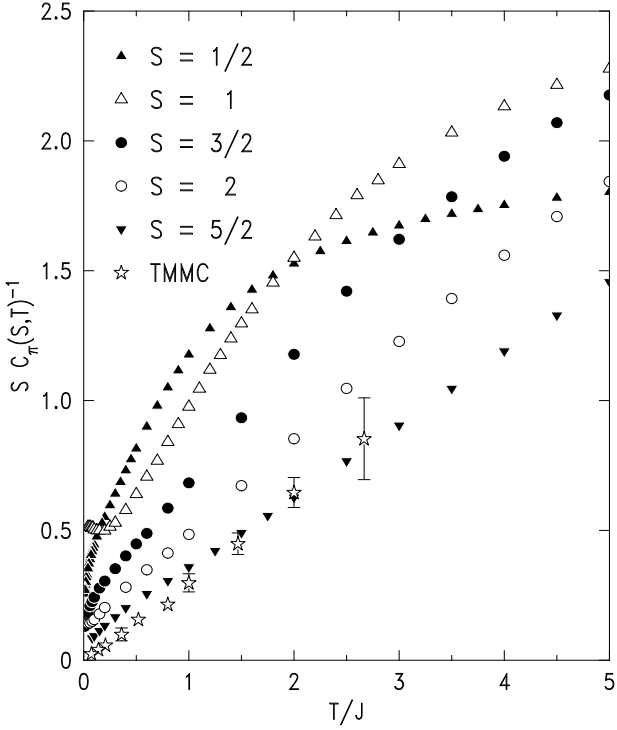}}
\caption{Inverse static structure factor peak intensity $C_\pi(S,T)^{-1}$,
multiplied by $S$ for graphical purposes, versus $T/J$. Note the deviation of
the quantum Monte Carlo results for $S=5/2$ from the data for TMMC around
$T/J \sim 1$.
}
\label{sst_TMMC}
\end{figure}

As shown in Fig.\ \ref{xi}, the correlation length in the spin-5/2 system 
TMMC is described well by the exact classical result. 
From Eq.\ 
(\ref{xi_WZNW}) one sees that quantum effects only modify the classical 
behavior $\xi \sim 1/T$ by a logarithmic correction factor. On the other 
hand, for $S=1/2$ the structure factor in Eq.\ (\ref{starykh_eq:b}) changes from the 
classical form $C_\pi 
(S,T) \sim 1/T$ to the quantum form $C_\pi (S,T) \sim [\ln(T_{\rm s} (S) / T)]^{3/2}$; 
this is a qualitative and not just a quantitative change due to quantum fluctuations in the 
divergence of $C_\pi (S,T)$ with decreasing temperature. 
Assuming this 
also holds for $S=3/2$ and $S=5/2$, we plot $C_\pi (S,T) ^{-1}$ versus $T$ in Fig.\ 
\ref{sst_TMMC}. It is evident that for $S=5/2$ there is a crossover from classical to 
quantum behavior around $T/J \sim 1$. Interestingly, the experimental data 
for TMMC do not seem to exhibit such a crossover. However, in TMMC there 
should also be a spin--space crossover from Heisenberg to XY behavior at 
the XY gap temperature, which is $\Delta_{\rm XY}/J \sim 0.7$. In the XY regime, $C_\pi 
(S,T)$ is enhanced. We speculate therefore that in TMMC the quantum and XY 
effects fortuitously cancel thus leading to the apparent classical $1/T$ 
behavior observed down to very low temperatures. Future quantum Monte Carlo 
calculations for $S=5/2$ including the XY anisotropy should serve to test
this conjecture.

\section{Discussion}
\label{sec:disc}

In a recent theoretical study of gapped spin chains at non--zero temperature, 
Damle and Sachdev\cite{Damle+Sachdev} developed a semiclassical picture based on thermally
excited particles for $T \ll \Delta$. In this
temperature regime, they were able to obtain expressions for
several dynamic properties of the $O(3)$ nonlinear $\sigma$ model. 
They also developed a rather different semiclassical approach for
$\Delta < T < J$ which is based upon classical waves described by the 
continuum $O(3)$ $\sigma$ model. Using this approach, Damle and Sachdev derive
one--loop expressions for the uniform susceptibility and correlation 
length in the temperature range $\Delta < T < J$:
\begin{equation}
\chi_{\rm u}(S,T)={1 \over {3\pi v_S}}
\left[
\ln\Bigl({{32 \pi T} \over {e^{\gamma + 2}\Delta_S}} \Bigr) 
+ \ln\ln\Bigl({{8 T} \over {e \Delta_S}} \Bigr)
\right]
\label{chi_Damle}
\end{equation}
\begin{equation}
\xi(S,T)={v_S \over {2 \pi T}}                   
\left[
\ln\Bigl({{32 \pi T} \over {e^{\gamma + 1}\Delta_S}} \Bigr) 
+ \ln\ln\Bigl({{8 T} \over {e \Delta_S}} \Bigr)
\right],
\label{xi_Damle}
\end{equation}
where $\gamma=0.5772...$ is Euler's constant.

\begin{figure}
\centerline{\epsfxsize=3.1in\epsfbox
{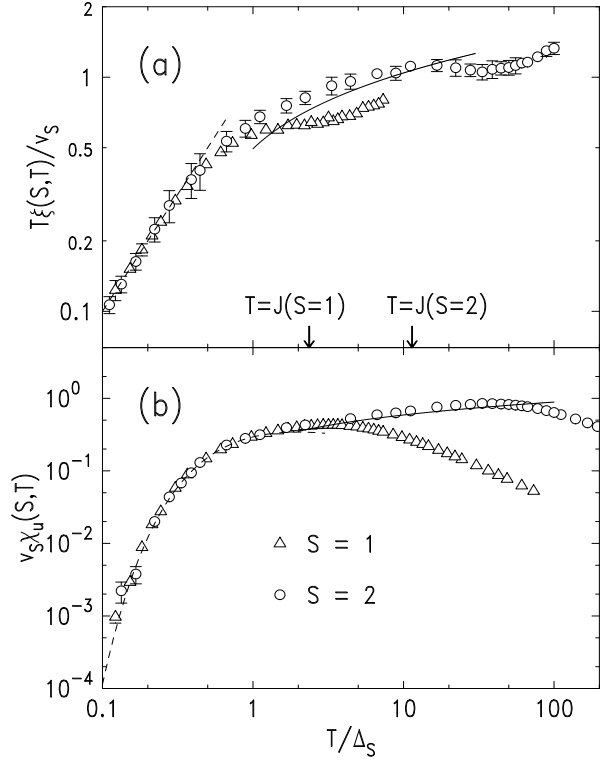}}
\caption{(a) Correlation length and (b) uniform susceptibility as a function of 
temperature for integer spin chains. The solid lines are the results Eqs.\ 
(\ref{chi_Damle}) and (\ref{xi_Damle}) without any adjustable parameter. 
The dashed line in (a) is the relation $v=\Delta\xi$,
and the dashed line in (b) is Eq.\ (\ref{chi_Tsvelik}). 
}
\label{damle} 
\end{figure}

In Fig.\ \ref{damle}, we show $T\xi(S,T)/v_S$ and $v_S \chi_{\rm u} (S,T)$ as
functions of $T/\Delta_S$ without any adjustable parameters, since 
$\Delta_S$ and $v_S$ have been determined independently.
For the $S=1$ chain we use values $\Delta_{S=1}=0.41050(2)$,
$\xi(1,0)/a=6.03(1)$, and $v_{S=1}/Ja=2.49(1)$ 
taken from the literature\cite{Sorensen,White}, since these
zero--temperature results have smaller error bars than our own values. 
The solid lines are Eqs.\ (\ref{chi_Damle}) and (\ref{xi_Damle}), which agree 
well with the $S=2$ data for $\Delta_{S=2} < T <J$. An even better agreement might be 
obtained if the theory were extended beyond the one--loop approximation.
It is evident, nonetheless, that a window of
temperature in which Eqs.\ (\ref{chi_Damle}) and (\ref{xi_Damle}) apply 
indeed exists for the $S=2$ chain, as speculated by Damle and 
Sachdev\cite{Damle+Sachdev}.

In Fig.\ \ref{damle}(b), we also show the scaling of the uniform 
susceptibility  at low 
temperatures ($T \ll \Delta$) along with the theoretical expression Eq.\ 
(\ref{chi_Tsvelik}) as a dashed line without any adjustable parameters. 
The dashed line in Fig.\ \ref{damle}(a) shows that the relation $v=\Delta\xi$ 
holds up to $T \approx \Delta/2$ in integer spin chains.

\begin{figure}
\centerline{\epsfxsize=3.3in\epsfbox
{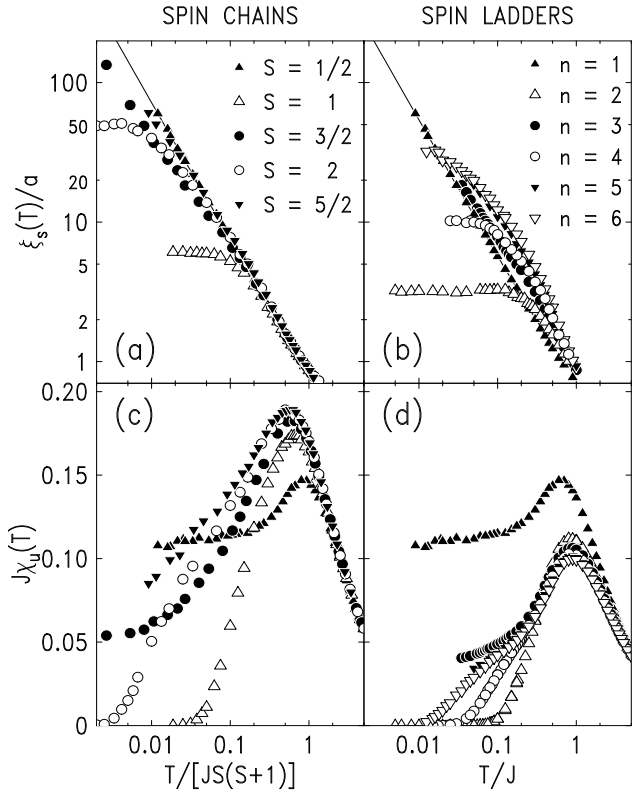}}
\caption{Uniform susceptibility per spin and staggered correlation length of 
the quantum spin chains are 
compared with those of quantum spin ladders, consisting of $n$ isotropically 
coupled antiferromagnetic $S=1/2$ chains\cite{Greven}: (a) correlation 
length of spin chains; 
(b) correlation length of spin ladders; (c) uniform susceptibility of 
spin chains; and (d) uniform susceptibility of spin ladders. Solid lines in (a) and (b) 
are the WZNW model prediction, Eq.\ \ref{xi_WZNW}} 
\label{comp}
\end{figure}

Spin-1/2 ladders are arrays of  
$n$ coupled Heisenberg chains with $S=1/2$.
In analogy to integer spin chains, ladders with an even number $n$
of chains exhibit exponentially decaying correlations in their ground 
state due to 
the presence of a spin gap, while those with odd $n$ show behavior similar 
to that of half--odd--integer spin chains. 
Therefore, it is illuminating to
plot the results such that the behaviors of quantum spin chains can be 
qualitatively compared 
with those of $S=1/2$ quantum spin ladders. In Fig.\ \ref{comp}, we display 
the staggered
correlation length and the uniform susceptibility per spin for both spin chains 
and isotropically coupled spin ladders\cite{Greven}. 
The same symbol is used for the spin--$S$ chain and the spin-1/2 
ladder of width $n=2S$. One observes the markedly different behaviors of 
gapped systems shown in open 
symbols compared with those of gapless systems shown as solid symbols.

In summary, antiferromagnetic Heisenberg spin chains with spin values ranging
from $S=1/2$ to $S=5/2$ have been studied with the quantum Monte Carlo method.
The temperature dependences of the uniform susceptibility, the staggered 
susceptibility, the static structure factor peak intensity,
and the correlation length are obtained. We find that at
high temperatures these quantities
agree very well with the exact results for the classical spin chain, and that quantum 
effects become progressively more important as the temperature is decreased. 
In addition, our data for $\chi_{\rm u}$ and $\xi$ for the $S=2$ chain
in the intermediate temperature range $\Delta_{S=2} <  T < J$ are reasonably well 
predicted by the theory of Damle and Sachdev.

\begin{acknowledgement}
We would like thank S. Sachdev and R. R. P. Singh for valuable discussions.
This work was supported 
by the National Science Foundation--Low Temperature Physics Programs
under award number DMR 97--04532 and by the International Joint
Research Program of NEDO (New Economic Development Organization), Japan.
\end{acknowledgement}


\end{document}